\documentclass[aps,prl,twocolumn,showpacs,preprintnumbers,superscriptaddress]{revtex4}
\usepackage{bm,amsmath,amssymb,latexsym,mathrsfs,enumerate}
\usepackage[mathcal]{euscript}
\usepackage[hypertex]{hyperref}
\usepackage{graphicx}

\graphicspath{{figs/}}
\renewcommand{\vec}[1]{\bm{#1}}
\renewcommand{\url}[1]{}
\def\urlprefix{ }
\begin{document}

\title{Vortex polarity switching by a spin--polarized current}

\author{Jean--Guy Caputo}
    \email{caputo@insa-rouen.fr}
    \affiliation{Laboratoire de Math\'ematiques, INSA de Rouen, B.P. 8, 76131 Mont-Saint-Aignan cedex, France}
    \affiliation{Laboratoire de Physique theorique et modelisation, Universit\'e
    de Cergy-Pontoise and C.N.R.S., Cergy-Pontoise, France}

\author{Yuri Gaididei}
    \email{yugaid@uni-bayreuth.de}
    \affiliation{Institute for Theoretical Physics, 03143 Kiev, Ukraine}

\author{Franz G.~Mertens}
    \email{franz.mertens@uni-bayreuth.de}
    \affiliation{Physics Institute, University of Bayreuth, 95440 Bayreuth, Germany}

\author{Denis D. Sheka}
     \email{denis\_sheka@univ.kiev.ua}
     \affiliation{National Taras Shevchenko University of Kiev, 03127 Kiev, Ukraine}

\date{\today}

%
%

\begin{abstract}

The spin-transfer effect is investigated for the vortex state of a magnetic
nanodot. A spin current is shown to act similarly to an effective magnetic
field perpendicular to the nanodot. Then a vortex with magnetization
(polarity) parallel to the current polarization is energetically favorable.
Following a simple energy analysis and using direct spin--lattice simulations,
we predict the polarity switching of a vortex. For magnetic storage devices,
an electric current is more effective to switch the polarity of a vortex in a
nanodot than the magnetic field.

\end{abstract}




\maketitle

The control of magnetic nonlinear excitations (domain walls and vortices) via
an electric current is of special interest for applications in spintronics
\cite{Tserkovnyak05,Bader06}. The spin--transfer effect, theoretically
predicted by \citet{Slonczewski96} and \citet{Berger96} and experimentally
verified in \cite{Tsoi98,Myers99,Krivorotov05}, is very promising for
magnetization switching with possible applications in magnetic storage
devices. A perspective candidate for this program is the vortex (ground) state
of a disk--shaped nanoparticle (nanodot), which provides high density storage
and high speed magnetic RAM \cite{Cowburn02}.

In this Letter we apply the spin--transfer effect to a nanodot in the vortex
state. Using a continuum description for the magnetization dynamics we show
that a spin current mainly acts like an effective magnetic field, applied to
the sample along the $z$--axis. The influence of a perpendicular field on the
vortex properties was studied in Refs.~\cite{Ivanov95b,Ivanov02} for
easy--plane magnets. There it was shown that the vortex behavior depends on
its core magnetization (polarity).  A vortex with a magnetization pointing in
the field direction (light vortex) has less energy than a vortex with a
magnetization pointing away from the field direction (heavy vortex). For a
strong enough field the heavy vortex loses its stability \cite{Ivanov02} and
switches into a light vortex as confirmed by Monte Carlo simulations
\cite{Lee04}. The qualitative similarity between the current torque term and
the magnetic field suggests that an appropriate current can switch the
polarity of a vortex. Using a simple energy argument we show that the vortex
polarity can be \emph{irreversibly} switched by a spin--polarized current.
These analytical predictions are confirmed by lattice spin dynamics. Since an
electric current is easier to apply and less energetic than a magnetic field,
we show that this is a more effective way of flipping the polarity of a vortex
than the magnetic field.

\begin{figure}[h]
\includegraphics[width=\columnwidth]{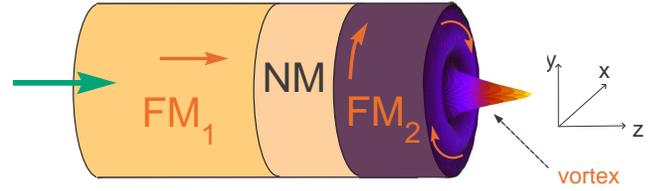}
\caption{(Color online). Schematic of the heterostructure.} \label{fig:hetero}
\end{figure}

To realize the program we use a pillar structure, first proposed in
Ref.~\cite{Kent04} (see Fig.~\ref{fig:hetero}). An electron current is
injected in $\text{FM}_1$ in $z$--direction, where it is polarized along $z$.
The second ferromagnetic layer $\text{FM}_2$ is separated from $\text{FM}_1$
by a narrow (about $2-3$ nm) nonmagnetic layer $\text{NM}$
\cite{Kent04,Xi05,Xi05b}, so the spin polarization of the current is conserved
when it flows into $\text{FM}_2$. A spin torque, acting on spins of the free
layer $\text{FM}_2$, causes their precession around $z$ with a constant
out-of--plane component \cite{Slonczewski96}.

The magnetization dynamics in the thin free layer $\text{FM}_2$ is described
by the Landau--Lifshitz--Gilbert equation with account of the spin--torque
effect by a Slonczewski--Berger term \cite{Slonczewski96,Berger96}. This
equation in dimensionless form is
\begin{equation} \label{eq:LLS-discrete}
\begin{split}
\frac{\mathrm{d} \vec{S}_{\vec{n}} }{\mathrm{d}t} =  &
\left[\vec{S}_{\vec{n}}\times \vec{F}_{\vec{n}}\right] - \varepsilon
\left[\vec{S}_{\vec{n}}
\times \frac{\mathrm{d} \vec{S}_{\vec{n}} }{\mathrm{d}t}\right]\\
& + j\,a(\eta)\,\Biggl[\vec{S}_{\vec{n}} \times \left[\vec{S}_{\vec{n}} \times
\frac{\vec{\sigma}}{1+b(\eta)\, \vec{S}_n\cdot \vec{\sigma}} \right] \Biggr],
\end{split}
\end{equation}
where  $\vec{S}_{\vec{n}}\equiv\left(S^x_{\vec{n}}, S^y_{\vec{n}},
S^z_{\vec{n}}\right)$ is a classical spin vector with fixed length $1$ on the
site $\vec{n}$ of the lattice, $\varepsilon$ is a damping coefficient. The
effective magnetic field is given by $\vec{F}_n=-\partial
\mathcal{H}/\partial\vec{S}_n$,  where $\mathcal{H}$ is the magnetic energy of
the free layer. We consider the  model of a classical easy-axis ferromagnet
described by the following Hamiltonian
\begin{equation} \label{eq:H-discrete}
\begin{split}
\mathcal{H} = &-\frac{1}{2}\!\! \sum_{\left(\vec{n},\vec{n}'\right)}\!
\left(\vec{S}_{\vec{n}}\cdot \vec{S}_{\vec{n}'} - \delta S^z_{\vec{n}}
S^z_{\vec{n}'} \right)
 +\mathcal{H}_{\text{dd}} .
\end{split}
\end{equation}
The summation runs over nearest--neighbor pairs $(\vec{n},\vec{n}')$,
$0<\delta\,<\, 1$ is the anisotropy constant,  ${\cal H}_{dd}$ is the magnetic
dipole-dipole interaction. The unit vector $\vec{\sigma}$ gives the direction
of the spin polarization ( along $z$ in our case), $j=J_e/J_p$ is the
normalized spin current, $J_e$ is the electrical current density, $J_p=\mu_0
M_S^2\,|e|\,d/\hbar$, where  $d$ is the disk thickness, $e$ is the electron
charge, $\mu_0$ is the vacuum permeability, $M_S$ is the saturation
magnetization. The functions $a(\eta)$ and $b(\eta)$ have the form
\cite{Slonczewski96}
\begin{equation*}
a(\eta)=\frac{4 \eta^{3/2}}{3 (1+\eta)^3-16\,\eta^{3/2}},\quad
b(\eta)=\frac{(1+\eta)^3}{3 (1+\eta)^3-16\,\eta^{3/2}},
\end{equation*}
where $0 < \eta < 1 $ is the degree of  spin polarization.

\begin{figure*}
\includegraphics[width=0.32\textwidth]{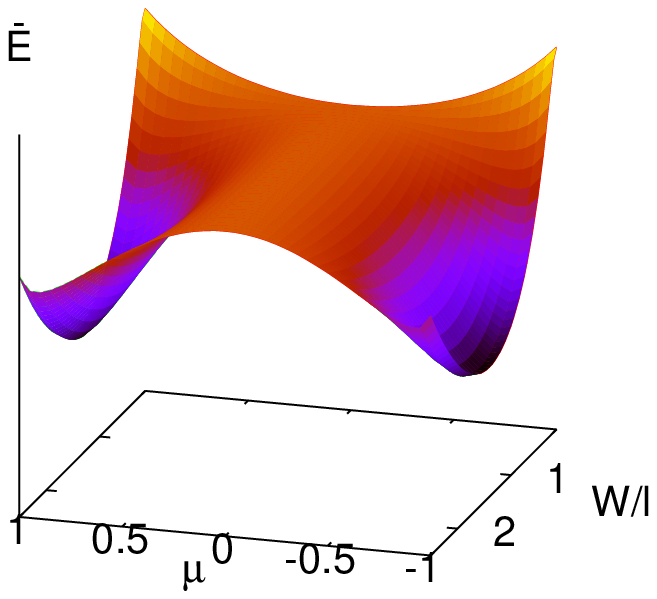}
\includegraphics[width=0.32\textwidth]{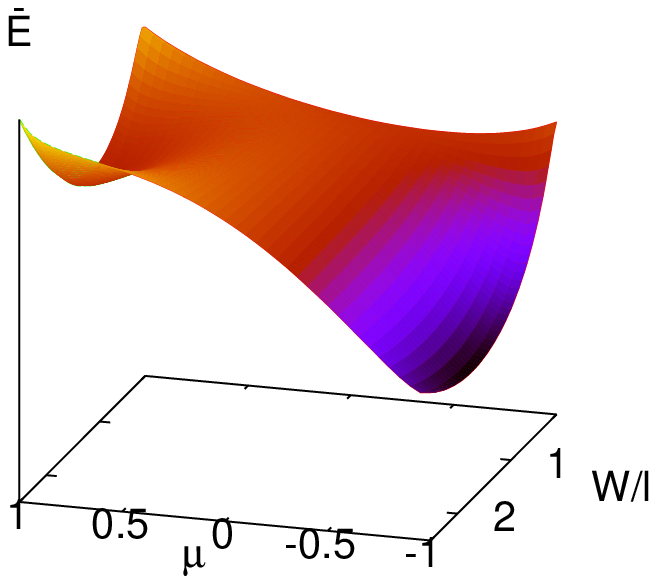}
\includegraphics[width=0.32\textwidth]{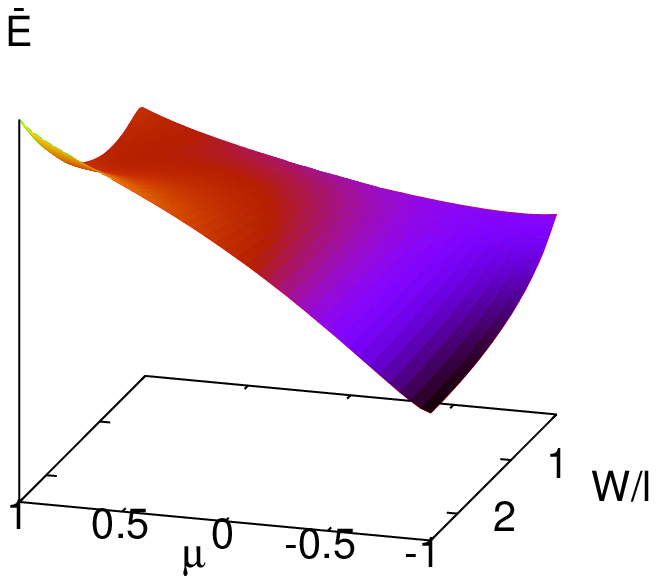}
\caption{(Color online). The energy of the vortex \eqref{eq:Energy-3} for
three different fields: $m_0=0$ (left panel), $m_0=-0.2$ (middle panel) and
$m_0=-0.5$ (right panel). The cutoff parameter $a=0.264 a_0$ and
$\ell = 2.89 a_0$. The switching occurs for $m_0\approx -0.47$.}%
\label{fig:Energy_h}
\end{figure*}

In weakly anisotropic magnets with $\delta \ll 1$, the characteristic size of
excitations $\ell_{\delta}=1/\sqrt{z\,\delta}$ is larger than the lattice
constant $a=1$ ($z$ is the number of nearest neighbors in the lattice), so
that one can use the continuum approximation for the Hamiltonian
\eqref{eq:H-discrete}
\begin{equation} \label{eq:H-cont}
\begin{split}
\mathscr{E} &= \mathscr{E}_0 +
\frac{1}{2}\,\int\,\mathrm{d}\vec{r}\Bigl[\left(\vec\nabla
\theta\right)^2 +\sin^2\theta\,\left(\vec\nabla\phi\right)^2\\
&+\frac{1}{
\ell_{\delta}^2}\,\cos^2\theta-\vec{H}_{\text{dm}}\cdot\vec{S}\Bigr],
\end{split}
\end{equation}
where $\mathscr{E}_0$ is a constant,  $\vec{H}_{\text{dm}}$ is an effective
demagnetization field which stems from the dipole--dipole interaction, and the
parameterization $\vec{S}=\left(\sin\theta\cos\phi, \sin\theta\sin\phi, \cos
\theta\right)$ is used. It is well known \cite{Usov93,Cowburn00} that a flat
cylindrical (disk--shaped) nanodot  with a radius $R$  smaller than a critical
value $R_0$ is in an in--plane mono--domain state, whereas for $R>R_0$ the dot
is in a vortex (curling) state.   The vortex state for $R>R_0$ becomes
energetically preferable as a result of the competition between the exchange
and the dipole--dipole interactions. Typical values are $R_0=50$ nm for a
permalloy nanodisk with $d=15$ nm \cite{Cowburn02}. Generally, the
demagnetization field is a nonlocal functional of $\vec{S}$, but for flat
dots, \emph{i.e.} for  dots with the aspect ratio $d/R\ll 1$, the action of
and  the magnetostatic field can be presented in a localized form
\cite{Gioia97,Ivanov02a}
\begin{equation}\label{eq:H-dm}
\vec{H}_{\text{dm}}=-\frac{1}{\ell^2_{e}}\, \cos\theta\, \vec{e}_z,
\end{equation}
where $\ell_{e}$ is the exchange length \cite{Skomski03}; hence the
demagnetization field  is equivalent to an effective easy--plane anisotropy
with a typical magnetic length scale
$\ell=\ell_{\delta}\,\ell_{e}/\sqrt{\ell_{\delta}^2+\ell_{e}^2}~$.

Thus in the continuum limit the current driven dynamics of flat nanodots  is
described by the set of equations
\begin{equation} \label{eq:LLS-cont}
\begin{split}
\sin\theta\,\partial_t\phi &= -\frac{\delta \mathscr{E}}{\delta \theta}\,
- \varepsilon\,\partial_t\theta,\\
- \sin\theta\,\partial_t\theta &=-\frac{\delta \mathscr{E}}{\delta \phi}\, -
\varepsilon\,\sin^2\theta\,\partial_t\phi + \frac{j\, a(\eta)\,
\sin^2\theta}{\sigma+ b(\eta)\,\,\cos\theta},
\end{split}
\end{equation}
where the energy functional $\mathscr{E}$ is given by Eq.~\eqref{eq:H-cont}
with the demagnetization field given by Eq.~\eqref{eq:H-dm} and $\sigma=\pm 1$
gives the sign of the spin polarization of the current.

Let us  discuss the stationary solutions of Eqs.~\eqref{eq:LLS-cont} without
and with current first for homogeneous magnetization and then for a vortex
state. In mono--domain nanodots for $j=0$  the magnetization is uniform and
lies in the $xy$--plane. Equation \eqref{eq:LLS-cont} shows that a current
causes a finite out--of--plane magnetization
\begin{equation*}
m_0 \equiv \cos\theta_0 =
\frac{\sigma}{2b}\left(\sqrt{1+\frac{4jab\ell^2}{\varepsilon}} -
1\right),\quad \omega \equiv \partial_t\phi = \frac{m_0}{\ell^2},
\end{equation*}
where the in plane component $\phi$ rotates around the $z$--axis with the
frequency $\omega$. Because $|m_0| <1$ this state exists for small $j$, it is
linearly stable for $|j\,a\,\ell^2-\varepsilon\,b|\,<\,\varepsilon $. Together
with this state there is always the fixed point $\theta_0 = 0$ (resp.
$\theta_0 =\pi$) which is stable for $j\,a\,\ell^2-\varepsilon\,b\,>
\,\varepsilon $ (resp. $j\,a\,\ell^2-\varepsilon\,b\,> -\varepsilon $).

In the vortex state the magnetization structure is determined by the
expressions
\begin{equation} \label{eq:vortex}
\cos\theta = p\,f(r), \qquad \phi = \pm\frac{\pi}{2} +\chi,
\end{equation}
where $r$ and $\chi$ are the polar coordinates in the $xy$-plane. The
localized function $f(r)$ describes the core of the vortex. In the core the
magnetization is nearly perpendicular to the disk--plane and the index $p=\pm
1$, which is termed \emph{polarity} \cite{Huber82}, determines which way the
out-of-plane structure is oriented. In nanodots for which the vortex
\eqref{eq:vortex} is the ground state a current leads to a stationary state
where the in--plane spin component rotates with the angular frequency
$\omega$. The in--plane and out--of--plane fields can be written as 
\begin{equation*}
\theta=\theta(r), \qquad \phi=\pm\,\frac{\pi}{2}+\chi+\omega\,t+\psi(r),
\end{equation*}
and from \eqref{eq:LLS-cont} they follow the equations
\begin{equation*} \label{eq:psi}
\begin{split}
&\nabla_r^2\theta +  \frac{\sin\theta}{\ell^2}\left(\cos\theta-m_0\right) -
\sin\theta\cos\theta
\left[\frac{1}{r^2} + (\partial_r\psi)^2\right]=0,\\
& \nabla_r^2\psi + 2\cot\theta\,\partial_r\theta \partial_r\psi +
\frac{\varepsilon b m_0}{\ell^2} \frac{\cos\theta-m_0}{\sigma + b\cos\theta} =
0,
\end{split}
\end{equation*}
where $\nabla_r^2=r^{-1}\partial_r(r\partial_r)$ .

Thus, under the action of the current, the azimuthal angle $\phi$ acquires an
additional dependence $\psi$ on the radial coordinate $r$ roughly proportional
to $\varepsilon b m_0/\ell^2$. However, for weak damping ($\varepsilon \ll 1$)
and for small spin polarization $(\eta\ll 1$) this additional dependence does
not change significantly the spatial dependence of the out--of--plane
component $\cos\theta$. In this limit, in the rotating frame of reference
where $\phi$ is replaced by $\widetilde{\phi}=\phi+\omega\,t$, the
Eqs.~\eqref{eq:LLS-cont} take on the usual Landau--Lifshitz--Gilbert form
without spin--torque term. A spin current in this case is equivalent to an
effective magnetic field applied along the hard $z$-axis. The energy of the
system in the rotating frame reads:
\begin{equation} \label{eq:Energy-j}
\widetilde{\mathscr{E}} = \mathscr{E}-\int\omega\cos\theta\,\mathrm{d}\vec{r}
\end{equation}
and the model under consideration is similar to the so--called cone--state
model \cite{Ivanov95b,Ivanov02}. It was shown in Refs.
\cite{Ivanov95b,Ivanov02} that  the magnetic field removes the degeneracy with
respect to the vortex polarity $p$. There  are light (heavy) vortices, where
the core is magnetized along (against) the field.  The light vortex has always
lower energy than the heavy one \cite{Ivanov95b}. The range of anisotropy and
magnetic field for which vortices are stable was found in
Ref.~\cite{Ivanov02}. The above mentioned qualitative similarity between the
current torque term and the magnetic field suggests   that by applying an
appropriate current one can achieve switching between the vortex states with
different polarities. It is worth reminding that in the continuum limit the
vortex states with different polarities are separated by an infinite barrier.
However, in  real  systems with their discrete lattice structure, the barrier
is finite \cite{Wysin94} and switching can occur (see \emph{e.g.} Refs.
\cite{Gaididei99,Gaididei00,Zagorodny03} where switching due to the noise or
an ac magnetic field was investigated).

To give some insight into  the current--induced switching mechanism, we
propose here a simple continuum picture, where the discreteness effects are
modeled by the cut--off parameter $a$, which is of the order of the lattice
constant $a_0$. We introduce the simple \emph{two--parameters Ansatz}
\begin{equation} \label{eq:new-Ansatz}
\begin{split}
\cos\theta &= \left(\mu - m_0\right)\, e^{-r^2/W^2} + m_0,\qquad \phi=\pm
\frac{\pi}{2}+\chi,
\end{split}
\end{equation}
where the variational parameter $\mu$ characterizes the out--of--plane core
magnetization while the parameter $W$ gives the core width. Using this ansatz,
one can rewrite the energy \eqref{eq:Energy-j} in the form
$\widetilde{\mathscr{E}} \approx \pi d \left[(1-m_0^2)\ln(L/a) +
\mathcal{E}\right]$, where
\begin{equation}\label{eq:Energy-3}
\begin{split}
\mathcal{E} &= 2\!\!\!\!\int\limits_0^{\exp(-\xi)} \!\!\!\! \frac{t\ln t\,
\mathrm{d}t}{\left(t-\frac{1-m_0}{\nu}\right)
\left(t+\frac{1+m_0}{\nu}\right)} - \nu m_0 \text{E}_1(\xi)\\
& - \frac{\nu^2}{2}\text{E}_1(2\xi) + \frac{\nu^2 W^2}{4\ell^2},\qquad \nu =
\mu - m_0,\; \xi = \frac{a^2}{W^2}.
\end{split}
\end{equation}
Here $\text{E}_1(\xi)$ is the exponential integral. Without the spin--current,
there exists a double--well potential, which provides equal minima for
$\mu=p=\pm1$ and $W=\ell\sqrt2$, see the left panel of
Fig.~\ref{fig:Energy_h}. Such minima exist only if the cutoff parameter
$a/\ell$ is smaller than some critical value which corresponds to the
transition between out--of--plane and in--plane vortices \cite{Wysin94}. For
the square lattice such a transition is realized for $\delta=0.297$
\cite{Wysin94} or $\ell\approx0.918a_0$; this corresponds to
$a/\ell\approx0.288$ and $a\approx 0.264 a_0$.

If the current is turned on, one of the vortex states, namely the vortex
polarized along $m_0$, becomes energetically preferable, see the middle panel
of Fig.~\ref{fig:Energy_h}. When the current exceeds some critical value, the
barrier disappears, and the switching occurs, see the right panel of
Fig.~\ref{fig:Energy_h}. In terms of this variational Ansatz, the switching
phenomenon can be considered as the motion of an effective mechanical particle
in the two-dimensional potential \eqref{eq:Energy-3}. For small currents the
particle is captured in one of the potential minima but when the current
exceeds a threshold value one of the minima disappears and the particle moves
to the another one. The vortex core width $W\approx
\ell\sqrt{2(p+m_0)/(p-m_0)}$ increases with the current for light vortices and
decreases for heavy ones, which agrees with results for the static field
\cite{Ivanov95b}. The critical value of $m_0$, where switching occurs, depends
on the value of a cut--off parameter, $a/\ell$: for the parameters presented
on Fig.~\ref{fig:Energy_h}, the critical value is $m_0\approx-0.47$.

To validate our analytical predictions, we performed spin--lattice
simulations. We stress that the standard micromagnetic simulations are
\emph{not adequate} for this problem, because the switching phenomenon has a
discrete nature, while micromagnetic simulations are a discretization of
originally continuum Landau--Lifshitz equations. We have integrated
numerically the discrete (spin--lattice) Eqs.~\eqref{eq:LLS-discrete} over
square lattices of size $L\times L$ using a 4th--order Runge--Kutta scheme
with time step $0.01$ and free boundary conditions. The spin system is defined
by a circular border with the diameter $L$. We have fixed an exchange
easy--plane anisotropy parameter $\delta=0.03$ instead of the dipolar term,
corresponding to $\ell/a_0 =2.89$, and we used the damping parameter
$\varepsilon=0.01$. We have considered system sizes in the range
$L/a_0\in(40,200)$.

\begin{figure}
\includegraphics[width=\columnwidth]{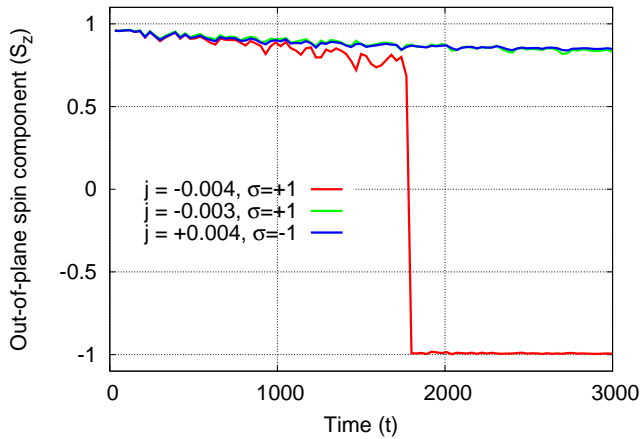}
\caption{(Color online). Amplitude of $S_z(t)$ at the vortex center from
numerical simulations on a square lattice with circular boundary of diameter
$L = 200$. Other parameters of the system: $\eta=0.25$, $\varepsilon=0.01$,
$\delta=0.03$; time is measured in units of $\hslash/J$. The critical current
$j\approx -0.0033$.}%
\label{fig:vortex}
\end{figure}

In our simulations we observed switching between heavy and light vortices as a
result of the influence of the current. Typical pictures are presented in
Fig.~\ref{fig:vortex}. The critical current $j\approx -0.0033$ is in a good
agreement with our continuum model calculations: the critical value of
$m_0\approx-0.47$ (see Fig.~\ref{fig:Energy_h}), which corresponds to
$j\approx-0.00332$.

To summarize, in this work the spin--transfer effect was investigated for the
vortex state nanodot. The switching of the vortex core magnetization was
studied analytically and confirmed numerically by direct spin--lattice
simulations. Let us estimate the critical current for a nanodot. Using typical
parameters for the permalloy disks \cite{Hoefer05} ($\eta=0.26$, $M_S=640$
kA/m, $\varepsilon =0.01$, $d\approx 10$ nm), one can estimate that
$J_{\text{cr}}\approx 10^{10} \text{A/m}^2$. The total current is about $10$
mA. For magnetic storage devices an electric current is then more effective to
switch the polarity of a vortex in a nanodot than the magnetic field which is
about the $0.1\div1$ T.

The authors thank S.~Demokritov (M\"unster University) for useful discussions.
Yu.G. and D.D.Sh. thank the University of Bayreuth, where part of this work
was performed, for kind hospitality and acknowledge the support from Deutsches
Zentrum f{\"u}r Luft- und Raumfart e.V., Internationales B{\"u}ro des BMBF in
the frame of a bilateral scientific cooperation between Ukraine and Germany,
project No.~UKR~05/055. J.G.C., Yu.G. and D.D.Sh. acknowledge support from a
Ukrainian--French Dnipro grant (No.~82/240293). D.D.Sh. acknowledges the
support from the Alexander von Humboldt--Foundation.


\end{document}